# Braid group action and quantum affine algebras

Jonathan Beck

Massachusetts Institute of Technology
Room 2-130
77 Massachusetts Avenue
Cambridge, MA 02139
email: beck@math.mit.edu

July 22, 1993

**Abstract.** We lift the lattice of translations in the extended affine Weyl group to a braid group action on the quantum affine algebra. This action fixes the Heisenberg subalgebra pointwise. Loop like generators are found for the algebra which satisfy the relations of Drinfel′d's new realization. Coproduct formulas are given and a PBW type basis is constructed.

**§0. Introduction.** The purpose of this paper is to explicitly establish the isomorphism between the quantum enveloping algebra $U_q(\widehat{\mathfrak{g}})$ of Drinfel′d and Jimbo ($\widehat{\mathfrak{g}}$ an untwisted affine Kac–Moody algebra) and the "new realization" [D2] of Drinfel′d. This is done using the braid group action defined on $U_q(\widehat{\mathfrak{g}})$ by Lusztig. In particular, we consider a group of operators $\mathcal{P}$ arising from the lattice of translations in the extended affine Weyl group.

Drinfel′d found that the study of finite dimensional representations of $U_q(\widehat{\mathfrak{g}})$ is made easier by the use of a "new realization" on a set of loop algebra–like generators over $\mathbb{C}[[h]]$. He gives (the proof is unpublished) an isomorphism to the usual presentation, although by his construction there is no explicit correspondence between the two sets of generators. Here we find the loop–like generators in $U_q(\widehat{\mathfrak{g}})$ and prove a version of [D2] which sits inside the Lusztig form over $\mathbb{Q}[q, q^{-1}]$. We also give formulas for the coproduct of the Drinfel′d generators.

The method is to show that $U_q(\widehat{\mathfrak{g}})$ contains $n$ (= rank $\mathfrak{g}$) "vertex" subalgebras $U_i$, each isomorphic to $U_q(\widehat{\mathfrak{sl}_2})$. Applying work of Damiani [Da], it follows that $U_q(\widehat{\mathfrak{g}})$ contains a Heisenberg subalgebra pointwise fixed by the group of translations $\mathcal{P}$. This subalgebra contains the purely imaginary Drinfel′d generators. We find the remaining generators as $\mathcal{P}$ translations of the usual Drinfel′d–Jimbo generators.

Having found expressions for imaginary root vectors in the usual presentation of $U_q(\widehat{\mathfrak{g}})$, it is a straightforward application to define a basis of Poincaré–Birkhoff–Witt type (with the method of [L5]).

**Acknowledgments.** I would like to thank Ian Grojnowski, Victor Kac and George Lusztig for helpful conversations.

**§1. Notation.** 1.1 We review the following standard notation (see [K]). Let $(a_{ij})$, $i, j \in I = \{0, \ldots, n\}$ be the $(n+1) \times (n+1)$ Cartan matrix of $\widehat{\mathfrak{g}}$ so that $(a_{ij})$, $1 \leq i, j \leq n$ is the Cartan matrix of the simple Lie algebra $\mathfrak{g}$. Let $d_i$ be relatively prime positive integers such that $(d_i a_{ij})$ is a symmetric matrix. Let $P^\vee$ be a lattice over $\mathbb{Z}$ with basis $\omega_i^\vee$, $1 \leq i \leq n$. Let

$\alpha_j^\vee = \sum_{i=1}^n a_{ji}\omega_i^\vee$, $1 \leq j \leq n$ and $Q^\vee = \sum_i \mathbb{Z}\alpha_i^\vee \subset P^\vee$. Then $P^\vee, Q^\vee$ are called respectively the coweight and coroot lattices of $\mathfrak{g}$. Let $Q_+^\vee = \sum_i \mathbb{Z}_+\alpha_i^\vee$, $P_+^\vee = \sum_i \mathbb{Z}_+\omega_i^\vee$.

Define the root lattice $Q = \text{Hom}(P^\vee, \mathbb{Z})$ with basis given by $\alpha_i$ such that $\langle \alpha_i, \omega_j^\vee \rangle = \delta_{ij}$. For $1 \leq i \leq n$ define the reflection $s_i$ acting on $P^\vee$ by $s_i(x) = x - \langle \alpha_i, x \rangle \alpha_i^\vee$. Additionally, $s_i$ acts on $Q$ by $s_i(y) = y - \langle y, \alpha_i^\vee \rangle \alpha_i$ for $y \in Q$. Let $W_0$ be the subgroup of $\text{Aut}(P^\vee)$ generated by $s_1, \ldots, s_n$. Let $\Pi = \{\alpha_1, \alpha_2, \ldots \alpha_n\}$, $\Pi^\vee = \{\alpha_1^\vee, \alpha_2^\vee, \ldots \alpha_n^\vee\}$. Define the root system (resp. coroot system) $R = W_0\Pi$ (resp. $R^\vee = W_0\Pi^\vee$), then the correspondence $\alpha_i \leftrightarrow \alpha_i^\vee$ extends to $R \leftrightarrow R^\vee$ and for $\alpha \in R$, $\langle \alpha, \alpha^\vee \rangle = 2$.

1.2 Using the $W_0$ action on $P^\vee$ define $W = W_0 \ltimes P^\vee$ where the product is given by $(s, x)(s', y) = (ss', {s'}^{-1}(x) + y)$. $P^\vee$ is characterized as the subgroup of $W$ consisting of elements with finitely many conjugates. For $s \in W_0$ write $s$ for $(s, 0)$. Similarly for $x \in P^\vee$ write $x$ for $(1, x)$.

Let $\theta$ be the highest root of $R$. Then writing $s_0$ for $(s_\theta, \theta^\vee)$, the set $\{s_0, \ldots, s_n\}$ generates a normal Coxeter subgroup $\tilde{W}$ of $W$ with defining relations determined by $(a_{ij})$. $\mathcal{T} = W/\tilde{W}$ is a finite group in correspondence with a certain subgroup of diagram automorphisms of the Dynkin diagram of $\hat{\mathfrak{g}}$ (see [B]). Identifying $\tau \in \mathcal{T}$ with such an automorphism, $\tau$ acts on $\tilde{W}$ by $\tau s_i \tau^{-1} = s_{\tau(i)}$, for $0 \leq i \leq n$. We have $W \cong \mathcal{T} \ltimes \tilde{W}$. The length function of $\tilde{W}$ extends to $W$ by setting $l_W(\tau w) = l_{\tilde{W}}(w)$, for $\tau \in \mathcal{T}, w \in \tilde{W}$. The semigroup $P_+^\vee$ has the properties:

$$l(s_i x) = l(x) + 1, \ 1 \leq i \leq n,$$
$$l(xy) = l(x) + l(y), \ x, y \in P_+^\vee.$$

Extend $Q$ to the affine root lattice $\tilde{Q} = \mathbb{Z}\alpha_0 \oplus Q$ and set $\delta = \alpha_0 + \theta$. Then $W$ acts as an affine transformation group on $\tilde{Q}$. In particular, for $x \in P^\vee$, $1 \leq j \leq n$, $x(\alpha_j) = \alpha_j - \langle \alpha_j, x \rangle \delta$. Introduce the symmetric bilinear form $(.|.): \tilde{Q} \times \tilde{Q} \to \mathbb{Z}$ determined by $(\alpha_i | \alpha_j) = d_i a_{ij}$.

Let $q_i = q^{d_i}$. Introduce the $q$–integer notation in $\mathbb{C}(q)$ by:

$$[n]_i = \frac{q_i^n - q_i^{-n}}{q_i - q_i^{-1}}, \quad [n]_i! = \prod_{k=1}^n [k]_i.$$

1.3 One defines the *quantum affine algebra* $U_q(\hat{\mathfrak{g}})(= U_q)$ of Drinfel'd and Jimbo as an algebra over $\mathbb{C}(q)$ on generators $E_i, F_i$ $(i \in I), K_\alpha$ $(\alpha \in \tilde{Q})$, $C^{\pm 1/2}, D^{\pm 1}$ subject to the following relations:

$$[K_\alpha, K_\beta] = [K_\alpha, D] = 0, \ K_\alpha K_\beta = K_{\alpha+\beta}, \ K_0 = 1,$$
$$C^{\pm 1/2} \text{ is central}, \ (C^{\pm 1/2})^2 = K_\delta^{\pm 1},$$
$$K_\alpha E_j K_\alpha^{-1} = q^{(\alpha|\alpha_j)} E_j, \ DE_j D^{-1} = q^{\delta_{0j}} E_j,$$
$$K_\alpha F_j K_\alpha^{-1} = q^{-(\alpha|\alpha_j)} F_j, \ DF_j D^{-1} = q^{-\delta_{0j}} F_j,$$
$$[E_i, F_j] = \delta_{ij} \frac{K_i - K_i^{-1}}{q_i - q_i^{-1}},$$
$$\sum_{s=0}^{1-a_{ij}} (-1)^s E_i^{(1-a_{ij}-s)} E_j E_i^{(s)} = 0, \ \sum_{s=0}^{1-a_{ij}} (-1)^s F_i^{(1-a_{ij}-s)} F_j F_i^{(s)} = 0.$$

Here $K_i = K_{\alpha_i}$ and $E_i^{(s)} = E_i/[s]_i!$. We have added the square root of the canonical central element $K_\delta$ for later notational convenience.

Introduce the $\mathbb{C}$–algebra automorphism $\Phi$, and anti-automorphism $\Omega$ of $U_q$, defined by:

$$\Phi(E_i) = F_i, \quad \Phi(F_i) = E_i, \quad \Phi(K_\alpha) = K_\alpha, \quad \Phi(D) = D, \quad \Phi(q) = q^{-1},$$
$$\Omega(E_i) = F_i, \quad \Omega(F_i) = E_i, \quad \Omega(K_\alpha) = K_{-\alpha}, \quad \Omega(D) = D^{-1}, \quad \Omega(q) = q^{-1},$$

As usual, let $U_q^+$ (resp. $U_q^-$) denote the span of monomials in $E_i$ (resp. $F_i$) and $T$ the span of monomials in $K_\alpha, C^{\pm 1/2}$ and $D^{\pm 1}$. Then $U_q = U_q^- \otimes T \otimes U_q^+$ [L, Ro]. $U_q^+$ is graded by $\tilde{Q}_+$ in the usual way and $U_q^+ = \oplus_\nu (U_q^+)_\nu$ where $\nu \in \tilde{Q}_+$. An element $x \in U_q^+$ is called homogeneous if $x \in (U_q^+)_\nu$ for some $\nu$. In this case let [c.f. L 1.1.1] $|x| = \nu$. Note that $|0| = \nu$ for all $\nu$. For $i \in I$ introduce the twisted derivations $r_i, {}_ir$ of $U_q^+$ [cf. L 1.2.13] defined uniquely as linear maps over $\mathbb{C}(q)$ with the properties:

$$r_i(1) = {}_ir(1) = 0, \; r_i(E_j) = {}_ir(E_j) = \delta_{ij},$$
$${}_ir(xy) = {}_ir(x)y + q^{(|x|,\alpha_i)} x \, {}_ir(y),$$
$$r_i(xy) = q^{(|y|,\alpha_i)} r_i(x) y + x r_i(y), \; x, y \text{ homogeneous}.$$

The Braid group $\mathcal{B}$ associated to $W$ is the group on generators $T_w$ ($w \in W$) with the relation $T_w T_{w'} = T_{ww'}$ if $l(w) + l(w') = l(ww')$. A reduced presentation of $w \in W$ is an expression $w = \tau s_{i_1} \ldots s_{i_n}$ where $l(w) = n$, $\tau \in \mathcal{T}$.

Recall that the braid group associated to $\tilde{W}$, whose canonical generators one denotes by $T_i = T_{s_i}, i \in I$, acts as a group of automorphisms of the algebra $U_q$ ([L]):

$$T_i E_i = -F_i K_i, \; T_i E_j = \sum_{s=0}^{-a_{ij}} (-1)^{s-a_{ij}} q_i^{-s} E_i^{(-a_{ij}-s)} E_j E_i^{(s)} \text{ if } i \neq j,$$
$$T_i F_i = -K_i^{-1} E_i, \; T_i F_j = \sum_{s=0}^{-a_{ij}} (-1)^{s-a_{ij}} q_i^{s} F_i^{(s)} F_j F_i^{(-a_{ij}-s)} \text{ if } i \neq j,$$
$$T_i K_\beta = K_{s_i \beta}, \quad \beta \in \tilde{Q}, \quad T_i(D) = D K_i^{-\delta_{i0}}.$$

Then $\Omega T_i = T_i \Omega$, and $\Phi T_i = T_i^{-1} \Phi$. We extend this action to $W$ by defining $T_\tau$ by $T_\tau(E_i) = E_{\tau(i)}, T_\tau(F_i) = F_{\tau(i)}, T_\tau(K_i) = K_{\tau(i)}$. Write $\tau$ for $T_\tau$. Denote by $\mathcal{P}$ the group generated by the operators $T_{\omega_i^\vee}$ ($1 \leq i \leq n$) and their inverses. From now on, for notational convenience refer to $\omega_i^\vee$ by $\omega_i$.

§**2. Some preliminary material.**

2.1 We review the following method (c.f. [DC–K], [L4], [L5]) of recovering the usual affine algebra through specialization at 1. Let $\mathcal{A}$ be the ring $\mathbb{C}[q, q^{-1}]$ localized at $(q-1)$. Let $U_\mathcal{A}$ be the $\mathcal{A}$ subalgebra of $U_q$ generated by the elements $E_i, F_i, K_i^{\pm 1}, D^{\pm 1}, C^{\pm 1/2}$, and:

$$H_i = \frac{K_i - K_i^{-1}}{q_i - q_i^{-1}}, \quad c = \frac{C - C^{-1}}{q - q^{-1}}, \quad d = \frac{D - D^{-1}}{q - q^{-1}}.$$

$U_\mathcal{A}$ includes the elements: $[K_i; d_i n] = H_i q_i^{-n} + K_i[n]_i$ and $[D; n] = dq^{-n} + D[n]$. Note the identities:
$$E_j H_i = [K_i; -d_i a_{ij}] E_j, \ F_j H_i = [K_i; d_i a_{ij}] F_j$$
$$E_0 d = [D; -1] E_0, \ F_0 d = [D; 1] F_0$$

Let $(q-1)U_\mathcal{A}$ be the left ideal generated by $(q-1)$ in $U_\mathcal{A}$. Define the algebra $\hat{U}_1$, the specialization of $U_q$ at 1, by $\hat{U}_1 = U_\mathcal{A}/(q-1)U_\mathcal{A}$. We obtain the following:

PROPOSITION [c.f. DC–K 1.5]. $\hat{U}_1$ is an associative algebra over $\mathbb{C}$ on the above generators with relations:

$c$ is central
$$[E_i, F_j] = \delta_{ij} H_i, [H_i, E_j] = a_{ij} K_i E_j, [H_i, F_j] = -a_{ij} K_i F_j,$$
$$[d, E_j] = \delta_{j0} D E_j, [d, F_j] = \delta_{j0} D F_j, K_i^2 = 1, D^2 = 1, C^2 = 1,$$
$$\mathrm{ad}^{(1-a_{ij})} E_i(E_j) = 0, \mathrm{ad}^{(1-a_{ij})} F_i(F_j) = 0, \ i \neq j.$$

In particular, $U_1 = \hat{U}_1/(K_i - 1, D - 1, C^{1/2} - 1)$ is isomorphic to the universal enveloping algebra of the affine Kac–Moody Lie algebra.

The following is due to Iwahori, Matsumoto and Tits.

PROPOSITION. Let $w \in W$ and let $\tau s_{i_1} s_{i_2} \ldots s_{i_n}$ be a reduced expression of $w$. Then the automorphism $T_w = \tau T_{i_1} T_{i_2} \ldots T_{i_n}$ of $U_q$ depends only $w$ and not on the reduced expression chosen. In particular, one reduced expression can be transformed to another by a finite sequence of braid relations.

We recall the following from [L]. The notation is adapted to this paper.

LEMMA [L 1.2.15]. Let $\nu \in \tilde{Q}_+, \nu \neq 0$. Let $x \in (U_q)_\nu$:
(a) If $r_i(x) = 0$ for all $i \in I$ then $x = 0$.
(b) If $_i r(x) = 0$ for all $i \in I$ then $x = 0$.

PROPOSITION [L 3.1.6]. Let $x \in U_q^+$, then:
$$[x, F_i] = \frac{r_i(x) K_i - K_i^{-1} {}_i r(x)}{q_i - q_i^{-1}},$$

PROPOSITION [L 38.1.6].
(a) $\{x \in U_q^+ | {}_i r(x) = 0\} = \{x \in U_q^+ | T_i(x) \in U_q^+\}$.
(b) $\{x \in U_q^+ | r_i(x) = 0\} = \{x \in U_q^+ | T_i^{-1}(x) \in U_q^+\}$.

PROPOSITION [L 40.1.2]. Let $w \in W, i \in I$ be such that $l(ws_i) = l(w)+1$. If $w = s_{i_1} \ldots s_{i_r}$ is a reduced presentation then $T_{i_1} \ldots T_{i_r}(E_i) \in U_q^+$.

LEMMA [L2 2.7]. Let $x \in P^\vee$, $i = 1\ldots n$, $s_i \in S$.
(a) If $s_i x = x s_i$ then $T_i T_x = T_x T_i$.
(b) If $s_i x s_i^{-1} = \alpha_i^{-1} x = \prod_j \omega_j^{a_j}$ then $T_i^{-1} T_x T_i^{-1} = \prod_j T_{\omega_j}^{a_j}$, in particular $T_i^{-1} T_{\omega_i} T_i^{-1} = T_{\omega_i}^{-1} \prod_{j \neq i} T_{\omega_j}^{-a_{ij}}$.

REMARK: $\Phi T_{\omega_i^{-1}} = T_{\omega_i}^{-1} \Phi$.

## §3. Subalgebras of $U_q(\widehat{\mathfrak{g}})$.

In this section we find certain subalgebras of $U_q(\widehat{\mathfrak{g}})$ which are isomorphic to $U_q(\widehat{\mathfrak{sl}_2})$.

3.1 LEMMA. *Let $\omega_i \in P^\vee$, $1 \leq i \leq n$.*
  (a) *Any reduced presentation of $\omega_i$ starts with $\tau s_j$ where $\tau \in \mathcal{T}$ and $\tau s_j = s_0 \tau$.*
  (b) *Any reduced presentation of $\omega_i$ ends with $s_i$.*

PROOF: For $j \neq 0$, $l(s_j \omega_i) = l(\omega_i) + 1$ and this implies (a). For (b), if $l(\omega_i s_j) < l(\omega_i)$ then $\omega_i(\alpha_j) < 0$ which is only the case when $i = j$.

DEFINITION. *For $1 \leq i \leq n$ let $\omega_i' = \omega_i s_i$. Then $l(\omega_i') = l(\omega_i) - 1$.*

REMARK. $T_{\omega_i'} = T_{\omega_i} T_i^{-1}$.

DEFINITION. *For $1 \leq i \leq n$ let $U_i \subset U_q(\widehat{\mathfrak{g}})$ be the subalgebra generated over $\mathbb{C}(q_i)$ by*
$$E_i,\ F_i,\ K_i^{\pm 1},\ T_{\omega_i'}(E_i),\ T_{\omega_i'}(F_i),\ T_{\omega_i'}(K_i^{\pm 1}),\ C^{\pm 1/2},\ D^{\pm d_i}.$$

It is clear $T_{\omega_i'} E_i \in U_q^+, T_{\omega_i'} F_i \in U_q^-$ since $l(\omega_i' s_i) = l(\omega_i) = l(\omega_i') + 1$.

The following is proved as in [L5 1.8]:

3.2 LEMMA. *Let $i, j \in I$ and let $w \in W$ be such that $w(\alpha_i) = \alpha_j$. Then $T_w(E_i) = E_j$.*

COROLLARY. *Let $1 \leq i \neq j \leq n$. Then for $x \in U_j$, $T_{\omega_i}(x) = x$.*

PROOF. $\omega_i(\alpha_j) = \alpha_j$.

DEFINITION. *For $1 \leq i \neq j \leq n$, $a_{ij} \leq 0$ introduce the elements:*
$$F_{ij} = -F_j F_i + q^{-(\alpha_i | \alpha_j)} F_i F_j,$$
$$E_{ij} = -E_i E_j + q^{(\alpha_i | \alpha_j)} E_j E_i.$$

3.3 LEMMA. *Let $1 \leq i \neq j \leq n$.*
  (a) $T_{\omega_i}(F_{ji}) = T_{\omega_j}(F_{ij})$,
  (b) $T_{\omega_i}(E_{ji}) = T_{\omega_j}(E_{ij})$.

PROOF: For (a) if $a_{ij} = 0$ then both sides of the equation equal 0. Otherwise, since the statement is symmetric in $i$ and $j$ we may assume $a_{ji} = -1$. Then:
$$T_{\omega_j}(F_{ij}) = T_{\omega_j}(T_j^{-1} F_i) = T_j T_{\omega_j}^{-1} T_{\omega_i}(F_i)$$
$$= T_{\omega_i} T_j(F_i) = T_{\omega_i}(F_{ji})$$

which implies (a). (b) follows by applying $\Omega$.

3.4 LEMMA. *Let $1 \leq i \leq n$, $[F_i, T_{\omega_i'}(E_i)] = 0$.*

PROOF: By [L 3.1.6, 38.1.6] it suffices to check that both $T_i T_{\omega_i'}(E_i) \in U_q^+$ and $T_i^{-1} T_{\omega_i'}(E_i) \in U_q^+$. Since $\omega_i \in P_+^\vee, l(s_i \omega_i' s_i) = l(\omega_i) + 1 = l(s_i \omega_i') + 1$ so that $T_i T_{\omega_i'}(E_i) \in U_q^+$. Now $T_i^{-1} T_{\omega_i'}(E_i) = T_i^{-1} T_{\omega_i} T_i^{-1}(E_i) = T_{\omega_i}^{-1}(E_i)$. Since $\Phi \circ \Omega(U_q^+) = U_q^+$ and $\Phi \circ \Omega(T_{\omega_i}^{-1}(E_i)) = T_{\omega_i^{-1}}(E_i)$ it is enough to check $T_{\omega_i^{-1}}(E_i) \in U_q^+$. This follows because $\omega_i \in P_+^\vee$ and $l(\omega_i^{-1} s_i) = l(\omega_i^{-1}) + 1$.

3.5 LEMMA. *Let* $1 \leq i \leq n$, $j \neq i, 0$, $[F_j, T_{\omega'_i}(E_i)] = -CK_i^{-1}T_{\omega_j}(F_{ij})$.

PROOF: $[F_j, T_{\omega'_i}(E_i)] = [F_j, T_{\omega_i}(-K_i^{-1}F_i)] = -T_{\omega_i}([F_j, K_i^{-1}F_i]) = -T_{\omega_i}(K_i^{-1})T_{\omega_i}(q^{-(\alpha_i|\alpha_j)}F_jF_i - F_iF_j) = -CK_i^{-1}T_{\omega_i}(F_{ji}) = -CK_i^{-1}T_{\omega_j}(F_{ij})$.

3.6 LEMMA. *Let* $1 \leq i \leq n$, *if* $\widehat{\mathfrak{g}} \neq \widehat{\mathfrak{sl}_2}$ *then* $r_0(T_{\omega'_i}(E_i)) = 0$.

PROOF: Since $l(\omega_i) > 1$ and $T_{\omega'_i}(E_i) \in U_q^+$ it follows $T_0^{-1}T_{\omega'_i}(E_i) \in U_q^+$.

3.7 PROPOSITION. *Let* $1 \leq i \leq n$.
   (a) $E_i^{(3)}T_{\omega'_i}(E_i) - E_i^{(2)}T_{\omega'_i}(E_i)E_i + E_iT_{\omega'_i}(E_i)E_i^{(2)} - T_{\omega'_i}(E_i)E_i^{(3)} = 0$,
   (b) $T_{\omega'_i}(E_i)^{(3)}(E_i) - T_{\omega'_i}(E_i)^{(2)}E_iT_{\omega'_i}(E_i) + T_{\omega'_i}(E_i)E_iT_{\omega'_i}(E_i)^{(2)} - E_iT_{\omega'_i}(E_i)^{(3)} = 0$.

PROOF: (b) follows from (a) by applying $T_{\omega'_i}$. Denote the expression in (a) by $x_i$. To check $x_i = 0$ it suffices [L 1.2.15] to check $r_j(x_i) = 0$ for $j \in I$. For $j = 0$ this is by the preceding lemma. Since for $x \in U_q^+$ $[F_j, x] = 0$ implies $r_j(x) = 0$, we can check that $[F_j, x_i] = 0$. This is straightforward using the expressions for $[F_j, T_{\omega'_i}(E_i)]$ in 3.4 and 3.5.

3.8 PROPOSITION. *For each* $1 \leq i \leq n$ *there is a algebra isomorphism* $h_i : U_q(\widehat{\mathfrak{sl}_2}) \to U_i$ *given by* $h_i(E_1) = E_i, h_i(E_0) = T_{\omega'_i}(E_i), h_i(K_1^{\pm 1}) = K_i^{\pm 1}, h_i(K_0^{\pm 1}) = T_{\omega'_i}(K_i^{\pm 1}), h_i(F_1) = F_i, h_i(F_0) = T_{\omega'_i}(F_i), h_i(C^{\pm 1/2}) = C^{\pm 1/2}, h_i(D^{\pm 1}) = D^{\pm d_i}, h_i(q) = q_i$.

PROOF: Consider the defining relations of $U_q(\widehat{\mathfrak{sl}_2})$. By the previous Proposition and some simple checks they hold in $U_i$ where $q$ is replaced by $q_i$. Therefore $h_i$ is surjective. For $\nu \in \tilde{Q}(\widehat{\mathfrak{g}})$, let $U_{i,\nu}^{\pm} = U_i \cap U_\nu^{\pm}$, then $h_{i|U_i^-}$ is homogeneous with respect to this grading. Therefore if $x \in \mathrm{Ker} h_{i|U_i^-}$, writing $x = \sum_j b_j x_j$ in terms of homogeneous components $h_{i|U_i^-}(x_j) = 0$ for each $j$. Fix some $x_j$. By [L4 Prop. 2.6] (see also remark 4.14) for $\beta \in \tilde{Q}$ there is a unique irreducible highest weight module $M$ of $U_q(\widehat{\mathfrak{sl}_2})$ with highest weight vector $v$ such that $K_iv = q^{(\alpha_i|\beta)}v$ for $i = 0,1$ and $Dv = q^{d_\beta}v$. Further we can pick $\beta$ so that $x_j$ acts non–trivially on $M$. The root system of $U_q(\widehat{\mathfrak{sl}_2})$ imbeds into that of $U_q(\widehat{\mathfrak{g}})$ via $h_i$ and we can fix a $\beta' \in \tilde{Q}(\widehat{\mathfrak{g}})$ so that pulling back the highest weight module $M'$ with weight $\beta'$ through $h_i$ we have $K_0, K_1$, and $D$ acting as on $M$. Now as a $U_q(\widehat{\mathfrak{sl}_2})$ module $M'$ has an irreducible quotient which is isomorphic to $M$. In particular, $x_j$ must act non–trivially in $M'$ which is a contradiction. Therefore Ker $h_{i|U_i^-} = 0$. Since multiplication induces a vector space isomorphism $U^- \otimes T \otimes U^+ \xrightarrow{m} U$ both in $U_i$ and $U_q(\widehat{\mathfrak{sl}_2})$ it follows that $h_i$ factors through this decomposition. Therefore Ker $h_i = 0$.

COROLLARY. *For* $1 \leq i \leq n$,
   (a) $T_{i|U_i} = h_i \circ T_1 \circ h_i^{-1}$,
   (b) $T_{\omega_i|U_i} = h_i \circ T_{\omega_1} \circ h_i^{-1}$.

PROOF: Let $M$ be an integrable $U_q$ module. Decompose $M$ into weight spaces with respect to the action of $K_i$, $M = \oplus_j M^j$. Let $u \in U_i$, $m \in M^n$ for a particular $n$. From the defining properties of the braid group action it follows:

$$T_1(h_i^{-1}(u)) \cdot \sum_{a,b,c;-a+b-c=n} (-1)^b q^{-ac+b} E_1^{(a)} F_1^{(b)} E_1^{(c)} m$$

$$= \sum_{a,b,c} (-1)^b q^{-ac+b} E_1^{(a)} F_1^{(b)} E_1^{(c)}(h_i^{-1}u)m$$

$$= h_i^{-1}\left(\sum_{a,b,c} (-1)^b q_i^{-ac+b} E_i^{(a)} F_i^{(b)} E_i^{(c)} um\right)$$

$$= h_i^{-1}\left(T_i(u) \sum_{a,b,c} (-1)^b q_i^{-ac+b} E_i^{(a)} F_i^{(b)} E_i^{(c)} m\right) \Longrightarrow h_i \circ T_1 \circ h_i^{-1} = T_{i|U_i}$$

This implies (a). $T_{\omega_i}^{-1}(E_i) = T_i^{-1} T_{\omega_i'}(E_i) = q_i^{-2} E_i^{(2)} T_{\omega_i'}(E_i) - q_i^{-1} E_i T_{\omega_i'}(E_i) E_i + T_{\omega_i'}(E_i) E_i^{(2)}$ and $T_{\omega_i}^{-1} T_{\omega_i'}(E_i) = -K_i^{-1} F_i$ so that $T_{\omega_i|U_i}$ acts on the generators of $U_i$ as does $h_i \circ T_{\omega_1} \circ h_i^{-1}$. (b) follows.

3.9 DEFINITION. For $1 \leq i \leq n$, $k > 0$, let $\overline{\psi}_{ik} = C^{-k/2}(q_i^{-2} E_i T_{\omega_i}^k(K_i^{-1} F_i) - T_{\omega_i}^k(K_i^{-1} F_i) E_i)$. Note that $\overline{\psi}_{ik} \in U_i$.

Versions of the next two propositions appear in the work of [Da §4] for $U_q(\widehat{\mathfrak{sl}_2})$.

3.10 PROPOSITION 1. Let $d = (q_i^2 C^{-1/2})$, $r > 0$, $m \in \mathbb{Z}$ then:

$$[\overline{\psi}_{ir}, T_{\omega_i}^m(F_i)] = -C^{1/2}[2]_i \left(\sum_{k=1}^{r-1} d^{(1-k)}(q_i - q_i^{-1}) \overline{\psi}_{i,r-k} T_{\omega_i}^{m+k}(F_i) + d^{(1-r)} T_{\omega_i}^{m+r}(F_i)\right)$$

$$[\overline{\psi}_{ir}, T_{\omega_i}^m(E_i)] = C^{-1/2}[2]_i \left(\sum_{k=1}^{r-1} d^{(k-1)}(q_i - q_i^{-1}) T_{\omega_i}^{m-k}(E_i) \overline{\psi}_{i,r-k} + d^{(r-1)} T_{\omega_i}^{m-r}(E_i)\right)$$

PROPOSITION 2. Let $r > 0$, $1 \leq i \leq n$.
  (a) $[\overline{\psi}_{i1}, \overline{\psi}_{ir}] = 0$,
  (b) $T_{\omega_i}(\overline{\psi}_{ir}) = \overline{\psi}_{ir}$.

PROOF: It is sufficient to prove the previous two statements for $U_q(\widehat{\mathfrak{sl}_2})$. Here $i = 1$ and $\omega_1 = \tau s_1$, where $\tau$ is the non–trivial Dynkin diagram automorphism. This follows because $l(\omega_1) = 1$, $l(s_1 \omega_1) = l(\omega_1) + 1$ and $\omega_1$ has only finitely many conjugates in $W$.

For the sake of exposition, we sketch a proof by induction on $r$ which appears in [Da §4]. For $r = 1$ the statements are readily checked. A direct calculation shows $[\overline{\psi}_{11}, \overline{\psi}_{1r}] = [2]((\tau T_1)^{-1}(\overline{\psi}_{1,r+1}) - \overline{\psi}_{1,r+1})$. This implies that 2a)$_r$ is equivalent to 2b)$_{r+1}$. Here we denote by 2a)$_{r'}$ the statement 2a) for all $r \leq r'$.

Proposition 2b)$_r$ implies 1)$_r$. This follows from an inductive calculation using the identities:

$$[\overline{\psi}_{1r}, F_1] = C^{1/2}\left(q^{-2}[\overline{\psi}_{1,r-1}, \tau T_1(F_1)] - [2](q - q^{-1}) \overline{\psi}_{1,r-1} \tau T_1(F_1)\right)$$
$$[\overline{\psi}_{1r}, E_1] = C^{-1/2}\left(q^2[\overline{\psi}_{1,r-1}, T_1^{-1}\tau(E_1)] + [2](q - q^{-1}) T_1^{-1}\tau(E_1) \overline{\psi}_{1,r-1}\right)$$

To show 2a) it is sufficient to show $r_j(C^{(r+1)/2}[\overline{\psi}_{11}, \overline{\psi}_{1r}]) = 0$ for $j = 0, 1$, $r > 0$. For $j = 0$ this is straightforward. For $j = 1$ this follows from $[[\overline{\psi}_{11}, \overline{\psi}_{1r}], F_1] = 0$. This is shown by induction on $r$. Assuming 2a)$_{r-1}$, 2b)$_r$, and 1)$_r$, a direct calculation gives

$$[[\overline{\psi}_{11}, \overline{\psi}_{1r}], F_1] = -C^{1/2}[2]\sum_{s=1}^{r-1} d^{(1-k)}[\overline{\psi}_{11}, \overline{\psi}_{1s}]T_{\omega_1}^k F_1 = 0.$$

This implies 2a)$_r$. As noted this now implies 2b)$_{r+1}$ and 1)$_{r+1}$. This completes the proof of Propositions 1 and 2.

REMARK: Much of the calculation through the end of §3 is inspired by the work of [Da] for $U_q(\widehat{sl}_2)$. The statements of Proposition 2 also appear for $U_q(\widehat{sl}_2)$ in [LSS].

3.11 Define $\overline{\varphi}_{ik} = \Omega(\overline{\psi}_{ik})$. Applying the anti–automorphism $\Omega$ to the above propositions gives similar identities with $\overline{\psi}_{ik}$ replaced by $\overline{\varphi}_{ik}$ and $F_i$ (resp. $E_i$) replaced by $E_i$ (resp. $F_i$). Here and in the future we omit writing these identities down although we implicitly assume them.

Let $H$ be the subalgebra of $U_q$ generated by $\overline{\psi}_{ik}, \overline{\varphi}_{ik}$ for $1 \leq i \leq n$, then we have shown:

3.12 PROPOSITION. *The group of translations $\mathcal{P}$ fixes $H$ pointwise.*

3.13 LEMMA. *Let $1 \leq i \leq n$, $r \in \mathbb{Z}$.*

$$T_{\omega_i}^r(F_i)F_i - q_i^{-2}F_i T_{\omega_i}^r(F_i) = q_i^{-2}T_{\omega_i}^{r-1}(F_i)T_{\omega_i}(F_i) - T_{\omega_i}(F_i)T_{\omega_i}^{r-1}(F_i).$$

PROOF: This is checked in $U_q(\widehat{sl}_2)$ directly.

3.14 LEMMA. *Let $a_{ij} \leq 0$, $m \in \mathbb{Z}$.*
  (a) $[\overline{\psi}_{i1}, T_{\omega_j}^m(F_j)] = C^{1/2}[a_{ij}]_i T_{\omega_j}^{m+1}(F_j),$
  (b) $[\overline{\psi}_{i1}, T_{\omega_j}^m(E_j)] = -C^{-1/2}[a_{ij}]_i T_{\omega_j}^{m-1}(E_j).$

PROOF: We check (a) for $a_{ij} \leq 0$. Note that by previous lemmas $[T_{\omega_i}(K_i^{-1}F_i), F_j] = -K_i^{-1}CT_{\omega_i}(F_{ji})$ and $T_{\omega_i}(F_{ji}) = T_{\omega_j}(F_{ij})$. Then:

$$[\overline{\psi}_{i1}, F_j] = C^{-1/2}([q_i^{-2}E_i T_{\omega_i}(K_i^{-1}F_i), F_j] - [T_{\omega_i}(K_i^{-1}F_i)E_i, F_j])$$
$$= -C^{1/2}K_i^{-1}([E_i, T_{\omega_i}(F_{ji})]) = -C^{1/2}K_i^{-1}T_{\omega_j}([E_i, F_{ij}])$$
$$= -C^{1/2}K_i^{-1}T_{\omega_j}([E_i, -F_j F_i + q^{-(\alpha_i|\alpha_j)}F_i F_j]) = C^{1/2}[a_{ij}]_i T_{\omega_j}(F_j)$$

Now (a) follows by applying $T_{\omega_j}^m$ to the above equality. Using $\psi_{i1} = T_{\omega_i}^{-1}\psi_{i1} = C^{-(1/2)}q_i^{-2}T_{\omega_i}^{-1}(E_i)(K_i^{-1}F_i) - (K_i^{-1}F_i)T_{\omega_i}^{-1}(E_i)$ (b) follows similarly.

3.15 LEMMA. *Let $a = a_{ij} \leq 0$, $r > 0$, $m \in \mathbb{Z}$, and let $d = (-q_i^a C^{-1/2})$.*

$$[\overline{\psi}_{ir}, T_{\omega_j}^m(F_j)] = C^{1/2}[a]_i \Big(\sum_{k=1}^{r-1} d^{(1-k)}(q_i - q_i^{-1})\overline{\psi}_{i,r-k}T_{\omega_j}^{m+k}(F_j) + d^{(1-r)}T_{\omega_j}^{m+r}(F_j)\Big),$$

$$[\overline{\psi}_{ir}, T_{\omega_j}^m(E_j)] = -C^{-1/2}[a]_i \Big(\sum_{k=1}^{r-1} d^{(k-1)}(q_i - q_i^{-1})T_{\omega_j}^{m-k}(E_j)\overline{\psi}_{i,r-k} + d^{(r-1)}T_{\omega_j}^{m-r}(E_j)\Big).$$

PROOF: We check the second equation.

$$[\overline{\psi}_{ir}, E_j] = C^{-r/2}(q_i^{-2}T_{\omega_i}^{-1}(E_i)T_{\omega_i}^{r-1}(K_i^{-1}F_i)E_j - T_{\omega_i}^{r-1}(K_i^{-1}F_i)T_{\omega_i}^{-1}(E_i)E_j$$
$$- q_i^{-2}E_j T_{\omega_i}^{-1}(E_i)T_{\omega_i}^{r-1}(K_i^{-1}F_i) + E_j T_{\omega_i}^{r-1}(K_i^{-1}F_i)T_{\omega_i}^{-1}(E_i))$$
$$\text{since:} \quad T_{\omega_i}^{r-1}(K_i^{-1}F_i)E_j = q_i^{-a} E_j T_{\omega_i}^{r-1}(K_i^{-1}F_i)$$
$$= C^{-r/2}(q_i^{-2-a}T_{\omega_i}^{-1}(E_i)E_j T_{\omega_i}^{r-1}(K_i^{-1}F_i) - T_{\omega_i}^{r-1}(K_i^{-1}F_i)T_{\omega_i}^{-1}(E_i)E_j$$
$$- q_i^{-2} E_j T_{\omega_i}^{-1}(E_i) T_{\omega_i}^{r-1}(K_i^{-1}F_i) + q_i^a T_{\omega_i}^{r-1}(K_i^{-1}F_i) E_j T_{\omega_i}^{-1}(E_i))$$
$$= C^{-r/2}\bigl(-q_i^{-2-a}T_{\omega_i}^{-1}(E_{ij})T_{\omega_i}^{r-1}(K_i^{-1}F_i) + T_{\omega_i}^{r-1}(K_i^{-1}F_i)T_{\omega_i}^{-1}(E_{ij})\bigr)$$
$$\text{now use:} \quad T_{\omega_i}^{-1}(E_{ij}) = T_{\omega_j}^{-1}(E_{ji}) = T_{\omega_j}^{-1}(-E_j E_i + q_i^a E_i E_j),$$
$$= C^{-r/2}(q_i^{-2-a}T_{\omega_j}^{-1}(E_j)E_i T_{\omega_i}^{r-1}(K_i^{-1}F_i) - q_i^{-2+a}E_i T_{\omega_i}^{r-1}(K_i^{-1}F_i)T_{\omega_j}^{-1}E_j$$
$$- q_i^{-a}T_{\omega_j}^{-1}(E_j)T_{\omega_i}^{r-1}(K_i^{-1}F_i)E_i + q_i^a T_{\omega_i}^{r-1}(K_i^{-1}F_i)E_i T_{\omega_j}^{-1}E_j)$$
$$= C^{-1/2}(-q_i^a[\overline{\psi}_{i,r-1}, T_{\omega_j}^{-1}(E_j)] - [a]_i(q_i - q_i^{-1})T_{\omega_j}^{-1}(E_j)\overline{\psi}_{i,r-1})$$

Now the second statement follows by induction and applying $T_{\omega_j}^m$. The first statement follows by a similar calculation.

3.16 LEMMA. Let $r \in \mathbb{Z}$, $(\alpha_i|\alpha_j) \leq 0$.

$$-T_{\omega_i}^r F_i F_j + q^{-(\alpha_i|\alpha_j)} F_j T_{\omega_i}^r F_i = q^{-(\alpha_i|\alpha_j)} T_{\omega_i}^{r-1} F_i T_{\omega_j} F_j - T_{\omega_j} F_j T_{\omega_i}^{r-1} F_i.$$

PROOF: The left hand side equals $T_{\omega_i}^r(F_{ji})$. The right hand side equals:

$$q^{-(\alpha_i|\alpha_j)} T_{\omega_i}^{r-1} F_i T_{\omega_j} F_j - T_{\omega_j} F_j T_{\omega_i}^{r-1} F_i = T_{\omega_i}^{r-1} T_{\omega_j}(F_{ij}) = T_{\omega_i}^r(F_{ji})$$

## §4. The relations in Drinfel'd's realization.

Let $\Gamma$ be the the Dynkin diagram of $\mathfrak{g}$. Orient the vertices of $\Gamma$ by defining $o : V \to \{\pm 1\}$ so that for $i$ and $j$ are adjacent in $\Gamma$, $o(i) = -o(j)$. Now define $\hat{T}_{\omega_i} = o(i)T_{\omega_i}$, and modify all the definitions by replacing $T_{\omega_i}$ with $\hat{T}_{\omega_i}$.

4.1 LEMMA. Let $a = a_{ij}$, $r > 0$, $m \in \mathbb{Z}$, and let $d = (q_i^a C^{-1/2})$.

$$[\overline{\psi}_{ir}, \hat{T}_{\omega_j}^m(F_j)] = -C^{1/2}[a]_i\bigl(\sum_{k=1}^{r-1} d^{(1-k)}(q_i - q_i^{-1})\overline{\psi}_{i,r-k}\hat{T}_{\omega_j}^{m+k}F_j + d^{(1-r)}\hat{T}_{\omega_j}^{m+r}F_j\bigr)$$

$$[\overline{\psi}_{ir}, \hat{T}_{\omega_j}^m(E_j)] = C^{-1/2}[a]_i\bigl(\sum_{k=1}^{r-1} d^{(k-1)}(q_i - q_i^{-1})\hat{T}_{\omega_j}^{m-k}E_j\overline{\psi}_{i,r-k} + d^{(r-1)}\hat{T}_{\omega_j}^{m-r}E_j\bigr)$$

PROOF: This follows directly from §3.

Now for $k > 0$ introduce generators $h_{ik} \in H$ by the change of variables (c.f. [D2], [G]):
$$(q_i - q_i^{-1})\sum_{k>0} h_{ik}z^k = \log(1 + (q_i - q_i^{-1})\sum_{k'>0} \overline{\psi}_{i,k'}z^{k'}),$$

Differentiating both sides and considering the coefficient of $z^r$ gives:

(\*) $$rh_{ir} = r\overline{\psi}_{ir} - (q_i - q_i^{-1})\sum_{k=1}^{r-1} k\overline{\psi}_{i,r-k}h_{ik}.$$

Similarly introduce $h_{i,-k} = \Omega(h_{ik})$ so that:

(\*\*) $$rh_{i,-r} = r\overline{\varphi}_{ir} - (q_i^{-1} - q_i)\sum_{k=1}^{r-1} kh_{i,-k}\overline{\varphi}_{i,r-k}.$$

4.2 LEMMA. *Let $1 \leq i, j \leq n$, $k > 0$.*
   (a) $[h_{ik}, \hat{T}^m_{\omega_j} F_j] = -\frac{1}{k}[ka_{ij}]_i C^{k/2} \hat{T}^{m+k}_{\omega_j} F_j$,
   (b) $[h_{ik}, \hat{T}^m_{\omega_j} E_j] = \frac{1}{k}[ka_{ij}]_i C^{-k/2} \hat{T}^{m-k}_{\omega_j} E_j$,

PROOF: Part (b) is an induction on $k$ using the following identities:
$$[\overline{\psi}_{ik}, \hat{T}^m_{\omega_j} E_j] = C^{-1/2}\big(-q_i^{a_{ij}}[\overline{\psi}_{i,k-1}, \hat{T}^{m-1}_{\omega_j} E_j] - [a_{ij}]_i(q_i - q_i^{-1})\hat{T}^{m-1}_{\omega_j} E_j \overline{\psi}_{i,k-1}\big),$$
$$[h_{ik'}, \hat{T}^m_{\omega_j} E_j] = C^{-1/2} \frac{(k'-1)[k'a_{ij}]_i}{k'[(k'-1)a_{ij}]_i}[h_{i,k'-1}, \hat{T}^{m-1}_{\omega_j} E_j], \text{ where } k' < k,$$
and (a) is similar.

REMARK. *As before, we omit the identities obtained by applying $\Omega$.*

For $1 \leq i \leq n$, $r > 0$, introduce the elements $\psi_{ir} = (q_i - q_i^{-1})K_i\overline{\psi}_{ir}$, $\varphi_{ir} = \Omega(\psi_{ir})$. Then:
$$\psi_{ir} = (q_i - q_i^{-1})C^{r/2}[E_i, \hat{T}^r_{\omega_i} F_i],$$
$$\varphi_{ir} = (q_i - q_i^{-1})C^{-r/2}[F_i, \hat{T}^r_{\omega_i} E_i].$$
Set $\psi_{i,0} = K_i$, $\varphi_{i,0} = K_i^{-1}$.

4.3 LEMMA. *Let $k, l \geq 1$. Then $[h_{ik}, \psi_{jl}] = 0$.*

PROOF:
$$\frac{1}{q_j - q_j^{-1}}[h_{ik}, \psi_{jl}] = C^{l/2}[h_{ik}, [E_j, \hat{T}^l_{\omega_j} F_j]]$$
$$= C^{l/2}[[h_{ik}, E_j], \hat{T}^l_{\omega_j} F_j] + [E_j, [h_{ik}, \hat{T}^l_{\omega_j} F_j]]$$
$$= C^{l/2}\Big(\frac{[ka_{ij}]_i}{k}C^{-k/2}[\hat{T}^{-k}_{\omega_j} E_j, \hat{T}^l_{\omega_j} F_j] + [E_j, -\frac{[ka_{ij}]_i}{k}C^{k/2}\hat{T}^{l+k}_{\omega_j} F_j]\Big)$$
$$= C^{k+l/2}\frac{[ka_{ij}]_i}{k}\big(C^{-k}[\hat{T}^{-k}_{\omega_j} E_j, \hat{T}^l_{\omega_j} F_j] - [E_j, \hat{T}^{l+k}_{\omega_j} F_j]\big) = 0,$$
$$\text{since } T^{-1}_{\omega_j}[E_j, T^l_{\omega_j} F_j] = C[E_j, T^l_{\omega_j} F_j].$$

Similarly:

4.4 LEMMA. *Let $k, r > 0$. Then*

$$[h_{ik}, \varphi_{jr}] = \begin{cases} -\dfrac{[ka_{ij}]_i}{k}(C^k - C^{-k})\varphi_{j,r-k} & \text{if } r \geq k, \\ 0 & \text{if } r < k. \end{cases}$$

Rewriting (**) in terms of the $\varphi_{ir}$ we have:

$$(***) \qquad r(q_i^{-1} - q_i)h_{i,-r} = rK_i\varphi_{ir} + (q_i - q_i^{-1})K_i \sum_{k=1}^{r-1} k\varphi_{i,r-k}h_{i,-k}.$$

4.5 LEMMA. *Let $k, l > 0$. Then*

$$[h_{ik}, h_{jl}] = \delta_{k,-l}\frac{1}{k}[ka_{ij}]_i\frac{C^k - C^{-k}}{q_j - q_j^{-1}}.$$

PROOF: Induction using (***).

4.6 DEFINITION. For $1 \leq i \leq n$, $k \in \mathbb{Z}$ define $x_{ik}^- = \hat{T}_{\omega_i}^k(F_i)$, $x_{ik}^+ = \hat{T}_{\omega_i}^{-k}(E_i)$.

We can now prove:

4.7 THEOREM [c.f. D2]. *$U_q(\hat{\mathfrak{g}})$ is generated over $\mathbb{C}(q)$ by the elements $x_{ij}^\pm, h_{ik}^\pm, K_i^{\pm 1}, C^{\pm 1/2}, D$, where $1 \leq i \leq n$, $j \in \mathbb{Z}$, and $k \in \mathbb{Z} \setminus \{0\}$. The following are defining relations for $U_q(\hat{\mathfrak{g}})$ :*

(1) $\quad [C^{\pm 1/2}, h_{ik}] = [C^{\pm 1/2}, x_{ik}^\pm] = [K_j, h_{ik}] = [K_i, K_j] = 0,$
$\quad K_i x_{jk}^\pm K_i^{-1} = q^{\pm(\alpha_i, \alpha_j)} x_{jk}^\pm, \quad Dx_{jk}^\pm D^{-1} = q^k x_{jk}^\pm, \quad Dh_{ik}D^{-1} = q^k h_{ik}$

(2) $\quad [h_{ik}, h_{jl}] = \delta_{k,-l}\dfrac{1}{k}[ka_{ij}]_i\dfrac{C^k - C^{-k}}{q_j - q_j^{-1}},$

(3) $\quad [h_{ik}, x_{jl}^\pm] = \pm\dfrac{1}{k}[ka_{ij}]_i C^{\mp(|k|/2)} x_{j,k+l}^\pm,$

(4) $\quad x_{i,k+1}^\pm x_{jl}^\pm - q^{\pm(\alpha_i|\alpha_j)} x_{jl}^\pm x_{i,k+1}^\pm = q^{\pm(\alpha_i|\alpha_j)} x_{ik}^\pm x_{j,l+1}^\pm - x_{j,l+1}^\pm x_{ik}^\pm,$

(5) $\quad [x_{ik}^+, x_{jl}^-] = \delta_{ij}\dfrac{1}{q_i - q_i^{-1}}\left(C^{k-l/2}\psi_{i,k+l} - C^{l-k/2}\varphi_{i,k+l}\right),$

For $i \neq j$, $n = 1 - a_{ij}$,

(6) $\quad \text{Sym}_{k_1,k_2,\ldots,k_n} \sum_{r=0}^{1-a_{ij}} (-1)^r \begin{bmatrix} n \\ r \end{bmatrix}_i x_{i,k_1}^\pm \ldots x_{i,k_r}^\pm x_{jl}^\pm x_{i,k_{r+1}}^\pm \ldots x_{i,k_n}^\pm = 0.$

Sym denotes symmetrization with respect to the indices $k_1, k_2, \ldots k_n$. Here $\psi_{ik}$ and $\varphi_{ik}$ are defined by the following functional equations:

$$\sum_{k\geq 0}\psi_{ik}u^k = K_i\exp\bigl((q_i - q_i^{-1})\sum_{k=1}^{\infty}h_{ik}u^k\bigr),$$

$$\sum_{k\geq 0}\varphi_{ik}u^k = K_i^{-1}\exp\bigl((q_i^{-1} - q_i)\sum_{k=1}^{\infty}h_{i,-k}u^{-k}\bigr).$$

PROOF: Relations (1)–(5) follow from the previous calculations. Relation (6) is obtained by applying $\hat{T}_{\omega_i}, i=1,\ldots,n$ to the Chevalley relations and an induction on $\max\{|k_{i_r} - k_{i_s}|\}$. Let $R$ be the algebra over $\mathbb{C}(q)$ on the above generators with defining relations (1)–(6). By the previous consideration there exists an algebra surjection $F: R \to U_q$. To check that $F$ is an isomorphism we specialize at 1 as in §2. Let $R_{\mathcal{A}}$ be the $\mathcal{A}$ subalgebra of $R$ generated by:

$$K_i^{\pm 1},\ C^{\pm 1/2},\ D^{\pm 1},\ h_{i,0} = \frac{K_i - K_i^{-1}}{q_i - q_i^{-1}},$$

$$c = \frac{C - C^{-1}}{q - q^{-1}},\ d = \frac{D - D^{-1}}{q - q^{-1}},\ h_{ik},\ x_{ik}^{\pm}$$

Define $\hat{R}_1 = R_{\mathcal{A}}/(q-1)R_{\mathcal{A}}$. Then $\hat{R}_1$ is an associative algebra over $\mathbb{C}$ on the above generators with the defining relations:

(1) $\quad [K_i, K_j] = [D, K_i] = 0,\ C^2 = D^2 = K_i^2 = 1,$
$\qquad [d, h_{ik}] = kh_{ik},\ [d, x_{jk}^{\pm}] = kDx_{jk}^{\pm},$

(2) $\quad [h_{ik}, h_{jl}] = \delta_{k,-l}a_{ij}\dfrac{c}{d_j}(C^{k-1} + \cdots + C^{1-k}),$

(3) $\quad [h_{ik}, x_{jl}^{\pm}] = \pm a_{ij}C^{\mp|k|/2}x_{j,k+l}^{\pm},\ [h_{i0}, x_{jl}^{\pm}] = \pm a_{ij}K_i x_{jl}^{\pm},$

(4) $\quad x_{i,k+1}^{\pm}x_{jl}^{\pm} - x_{jl}^{\pm}x_{i,k+1}^{\pm} = x_{ik}^{\pm}x_{j,l+1}^{\pm} - x_{j,l+1}^{\pm}x_{ik}^{\pm},$

(5) $\quad [x_{ik}^{+}, x_{jl}^{-}] = \delta_{ij}K_i C^{(k-l)/2}h_{i,k+l},$

(6) $\quad [x_{i,k_1}^{\pm}, [x_{i,k_2}^{\pm}, \ldots, [x_{i,k_n}^{\pm}, x_{jl}^{\pm}]\ldots] = 0,\quad n = 1 - a_{ij}.$

It follows from the Gabber–Kac theorem [G–K] (see [G] for the relations in $R_1$ below) that:

$$R_1 = \hat{R}_1/(K_i - 1, C^{1/2} - 1, D - 1) \cong U(\mathfrak{g} \otimes \mathbb{C}[t, t^{-1}] \oplus \mathbb{C}c \oplus \mathbb{C}d).$$

Now specialize $U_q(\widehat{\mathfrak{g}})$ to $U_1(\widehat{\mathfrak{g}})$ as in §2. Then $F$ induces the isomorphism:

$$F: R_1 \cong U_1(\widehat{\mathfrak{g}}),$$

Since specialization doesn't change the root multiplicities, $F: R \to U_q$ is an isomorphism.

REMARK: Let $s_{\theta_i} \in W_0$ so that $s_{\theta_i}(\alpha_i) = \theta$. By Lemma 3.2 it follows $T_{\theta_i}T_{\omega_i}(-K_i^{-1}F_i) = E_0$. This gives the inverse to the isomorphism $F: R \to U_q$. In particular, $F^{-1}(E_0) = -o(i)CK_\theta^{-1}T_{\theta_i}x_{i1}^{-}$.

## § 5. The coproduct.

Since the Drinfel'd generators are now expressed in terms of the braid group, calculating their coproduct depends on how the coproduct commutes with the braid group.

Define for $1 \le i \le n$:

$$R_i = \sum_{k \ge 0} (-1)^k q_i^{\frac{-k(k-1)}{2}} (q_i - q_i^{-1})^k [k]_i! T_i(F_i)^{(k)} \otimes T_i(E_i)^{(k)}$$

$$\overline{R}_i = (T_i^{-1} \otimes T_i^{-1}) R_i^{-1} = \sum_{k \ge 0} q_i^{\frac{k(k-1)}{2}} (q_i - q_i^{-1})^k [k]_i! F_i^{(k)} \otimes E_i^{(k)}.$$

The following proposition is due in the finite type case to [K–R], [L–S]. The Kac–Moody case is due to [L 37.3.2].

5.1 PROPOSITION. Let $S_i = T_i \otimes T_i$. Let $1 \le i \le n$, $x \in U_q$.
  (a) $\Delta(T_i(x)) = R_i^{-1} \cdot S_i \Delta(x) \cdot R_i$,
  (b) $\Delta(T_i^{-1}(x)) = \overline{R}_i^{-1} \cdot S_i^{-1} \Delta(x) \cdot \overline{R}_i$.

Let $\tau s_{i_1} \ldots s_{i_r}$ be a reduced presentation of $w$. Define

$$R_w = \tau \big( S_{i_1} S_{i_2} \ldots S_{i_{r-1}}(R_{i_r}) \ldots S_{i_1}(R_{i_2}) R_{i_1} \big),$$
$$\overline{R}_w = S_{i_r}^{-1} \ldots S_{i_2}^{-1}(\overline{R}_{i_1}) \ldots S_{i_r}^{-1}(\overline{R}_{i_{r-1}}) \overline{R}_{i_r}.$$

5.2 LEMMA. Let $w \in W$, $R_w$, $\overline{R}_w$ are well defined.

PROOF: If $W$ is the affine Weyl group of $\widehat{sl}_2$ any reduced presentation is unique. Otherwise, since any two reduced presentations differ by a finite sequence of braid relations it is enough to check the statement for the rank two case. Consider $R_{s_i s_j s_i}$, $R_{s_j s_i s_j}$ in the simply laced case. They are certainly equal since both (up to a torus element) are expressions for the rank 2 universal $R$–matrix (see [K–R], [L–S]).

5.3 PROPOSITION. Let $1 \le i \le n$, $k \ge 0$. Let $w = k\omega_i$.
  (a) $\Delta(x_{ik}^-) = R_w^{-1}(x_{ik}^- \otimes K_{-\alpha_i + k\delta} + 1 \otimes x_{ik}^-) R_w$,
  (b) $\Delta(x_{i,-k}^-) = \overline{R}_w^{-1}(x_{i,-k}^- \otimes K_{-\alpha_i - k\delta} + 1 \otimes x_{i,-k}^-) \overline{R}_w$.

PROOF: This follows inductively from the above formulas.

To obtain the coproduct on $x_{ik}^+$ note that $\Omega(x_{i,-k}^-) = x_{ik}^+$ and use $\Delta \circ \Omega = \Omega \otimes \Omega \circ \sigma \circ \Delta$ on the above formulas.

## §6. A PBW basis of $U_q$.

For $1 \le i \le n$, $C^{k/2} \overline{\psi}_{ik} \in U_{\mathcal{A}}^+$. On specialization to $q = 1$ these elements form a basis of the root space $k\delta$ of $\widehat{\mathfrak{g}}$. This follows from the previous section since $\overline{\psi}_{ik} = h_{ik} \mod (q-1)$, which implies their linear independence on specialization. Note that if $w(\alpha_i) = \beta$ ($\alpha_i$ simple, $\beta$ positive, $w \in W$) then $T_w(E_i)$ specializes to a root vector of $\widehat{\mathfrak{g}}$ of root $\beta$.

For $\beta \in \Delta_+^{\text{re}}(\widehat{\mathfrak{g}})$ choose $w_\beta \in \tilde{W}$ so that $w_\beta(\alpha_{i_\beta}) = \beta$ for some $i_\beta \in I$. Define $E_\beta = T_{w_\beta}(E_{i_\beta})$. For $\kappa : \Delta_+^{\text{re}} \to \mathbb{N}$, $\iota : \{1, \ldots, n\} \times \Delta_+^{\text{im}} \to \mathbb{N}$ define

$$E^{\kappa,\iota} = \prod E_\beta^{\kappa(\beta)} (C^{k/2} \overline{\psi}_{ik})^{\iota(i,k\delta)}, \ F^{\kappa',\iota'} = \Omega(E^{\kappa',\iota'})$$

where the product is in a predetermined total order over the positive roots counted with multiplicity.

6.1 PROPOSITION. *The $E^{\kappa,\iota}$ form a basis of $U_q^+(\widehat{\mathfrak{g}})$ as a $\mathbb{C}(q)$–vector space. The elements $F^{\kappa',\iota'} K_\alpha C^{r'+1/2} D^r E^{\kappa,\iota}$ ($\alpha \in \tilde{Q}$, $r, r' \in \mathbb{Z}, \kappa$, $\iota$ as above) form a basis of $U_q(\widehat{\mathfrak{g}})$ as a $\mathbb{C}(q)$–vector space.*

PROOF: The proof can be repeated almost word for word as found in [L5 §1]. In the proof of linear independence of the $E^{\kappa,\iota}$, a dominant integral highest weight should be chosen so that for $\kappa, \iota \in \mathfrak{G}$ (in the notation found there) the $\overline{E}^{\kappa,\iota}$ form a linearly independent set in $\overline{M}$.

REMARK: The above basis is called of Poincaré–Birkhoff–Witt type because on specialization to 1 it degenerates to a PBW basis of the enveloping algebra $U(\widehat{\mathfrak{g}})$.